# Early School Leaving in Spain: a longitudinal analysis by gender

Martín Martín-González[a], Sara M. González-Betancor[b] and Carmen Pérez-Esparrells[c]

[a] Universidad Europea de Canarias, Spain; [b] Universidad de Las Palmas de Gran Canaria, Spain; [c] Universidad Autónoma de Madrid, Spain and UAM-UC3M Research Institute for Higher Education and Science (INAECU)

# Early School Leaving in Spain: a longitudinal analysis by gender

Martín Martín-González[a], Sara M. González-Betancor[b] and Carmen Pérez-Esparrells[c]

[a] Universidad Europea de Canarias, Spain; [b] Universidad de Las Palmas de Gran Canaria, Spain; [c] Universidad Autónoma de Madrid, Spain and UAM-UC3M Research Institute for Higher Education and Science (INAECU)

**Abstract**

Spain is one of the eight EU-27 countries that failed to reduce early school leaving (ESL) below 10% in 2020, and now faces the challenge of achieving a rate below 9% by 2030. The determinants of this phenomenon are usually studied using cross-sectional data at the micro-level and without differentiation by gender. In this study, we analyse it for the first time for Spain using panel data (between 2002-2020), taking into account the high regional inequalities at the macroeconomic level and the masculinisation of the phenomenon. The results show a positive relationship between ESL and socioeconomic variables such as the adolescent fertility rate, immigration, unemployment or the weight of the industrial and construction sectors in the regional economy, with significant gender differences that invite us to discuss educational policies. Surprisingly, youth unemployment has only small but significant impact on female ESL.

**Keywords**

Early school leaving, compulsory education, gender, immigration, fertility

**Key messages**

- The early school leaving (ESL) rate in Spain exhibited a downward trend during the first two decades of the 21st century, yet remained above the EU target.
- ESL in Spain shows large regional disparities and has a markedly male-dominated character.
- The main macroeconomic factors influencing the incidence of ESL in Spain, with gender-specific nuances, include population density, fertility rates, the composition of the immigrant population, and the sectoral composition of the labour market.
- The reduction of ESL can be achieved by implementing improvements to sex education and careers guidance, as well as by reducing youth unemployment.
- To develop targeted interventions for further reducing ESL in Spain, it's essential to analyse microeconomic factors.

1. **Introduction**

Early school leavers are individuals aged 18-24 with only lower secondary education (i.e., ISCED 2) or less and who are not in education or training. The EU definition of early school leavers includes those who have not completed compulsory education or, having completed it, have not followed further education or training to upper secondary level or its equivalent. This definition is in line with the European Labour Force Survey. The exclusion of the under 18s avoids misclassification of students who are still studying during the survey[1].

Early school leaving (ESL) is associated with significant negative impacts in developed economies. Some authors link it to an increased likelihood of unemployment, living in poverty, low-skilled employment or lower wages, as well as problems of poor health, depression or crime (Bridgeland et al, 2006; Oreopoulos, 2007; Dickson and Harmon, 2011). There are significant costs associated with these situations, not only to the individuals affected, but also to society as a whole (Belfield, 2008). At the macroeconomic level, reducing ESL is crucial for improving a country's economic growth and social cohesion —given that formal education is the primary means of acquiring the skills demanded by the labour market and that increase productivity— (Choi & Calero, 2018) and particularly relevant in the context of a knowledge-based economy.

In the European context, it is particularly relevant to analyse the case of Spain, given the evolution of this indicator over the last decades and the challenge it currently faces. ESL in Spain has traditionally been the highest in the European Union. However, it has been progressively reduced, with more intensity in recent years. It has fallen from 40.4% in 1992 to 16% in 2020, closing gap with the EU-28 average (National Statistics Institute, INE, 2022c). Nevertheless, by the end of 2020, Spain was one of the eight countries[2] that did not meet the European 2020 Strategy's target of reducing ESL to 10%, with a gap of 6 percentage points - only behind Malta. The new European Council Resolution on the strategic framework for cooperation in education and training in the European Higher Education Area (EHEA) recently set a new target for this indicator: "The share of early leavers from education and training should be less than 9%, by 2030" (Council of the European Union, 2021). Thus, Spain faces a new challenge for 2030 and understanding the determinants of ESL in Spain is of paramount importance in order to implement appropriate educational and other socioeconomic policies.

---

[1] For additional details, refer to https://ec.europa.eu/commission/presscorner/detail/en/MEMO_11_52.
[2] The eight countries are Malta (16,7%), Spain (16%), Romania (15,6%), Italy (13,1%), Bulgaria (12,8%), Hungary (12,1%), Cyprus (11,5%) and Germany (10,1%).

This paper seeks to answer a fundamental question: How have macro socioeconomic conditions influenced early school leaving (ESL) in Spain over the last two decades? Our unique approach spans the first two decades of the 21st century, using panel models to go beyond simple correlations of one-year cross-sectional data regressions. We examine various characteristics of Spanish regions that are often overlooked, and present gender-specific results that add a unique dimension to our research. We also analyse different population characteristics, providing a novel macro perspective on this critical issue.

To achieve this objective the rest of this paper is structured as follows: section 2 begins with a review of the literature; section 3 introduces the Spanish context; section 4 presents the methodology and the descriptive and exploratory analysis of the data; section 5 analyses and discusses the results of the econometric model; finally, section 6 summarises the main conclusions of the paper which can be extrapolated to other countries with high levels of ESL as well as some recommendations.

## 2. Literature Review

The determinants of ESL are multiple, covering a wide spectrum, from macroeconomic aspects to psychological factors of the students, demographic and social phenomena, family and neighbourhood contexts, school characteristics, teaching methodologies, etc. In this paper we use a macroeconomic perspective, which allows us to analyse ESL in a comprehensive way, as an institutional phenomenon closely linked to the formal and regulatory aspects of the education system (such as educational expenditure decisions and their distribution between public and private schools), to the characteristics of the labour market (and its composition by sector) and to the characteristics of the population. However, while some of the determinants are empirically demonstrated, there are other variables for which empirical evidence is limited or non-existent.

A variable that is usually pointed out when considering the determinants of ESL is the expenditure on non-university education in different regions. This variable can influence school investment, student-teacher ratios and class size, variables that policymakers in Western countries use as proxies for educational quality. However, there is no concrete evidence in one direction in the international literature on the effects of increased public investment in education on the phenomenon of ESL. Hanushek (2003), for example, questions the impact of additional resources on educational outcomes. But Nozaki and Matsuura (2017) suggest paying closer attention to public school investment, as they found that public school spending at elementary schools is correlated with student academic skills in Japan. In the case of Spain, authors such as Mora et al (2010) conclude that there is a relationship between higher educational expenditure per student and lower ESL, both for males and females. Author's own, (2012) also found that the higher the public expenditure per student, the higher the percentage of graduates, which is associated with lower ESL. They conclude that public investment in secondary education is more relevant, but public expenditure in primary and early childhood education also affects the ESL rate. Bayón-Calvo et al (2021) also insist on considering all educational stages when analysing ESL, although focusing on endogenous determinants related to scholastic performance. They confirm the relationship between ESL and weak scholastic performance and, although they analyse regional differences, they cannot establish a direct relationship between educational expenditure and ESL.

On the other hand, a less studied variable is public expenditure as a function of school ownership (public, charter or private schools). In most micro-level studies in Spain, school ownership is no longer a relevant determinant of academic performance once the socioeconomic environment of schools is considered (Choi & Calero, 2018). In the case of Spain, it is interesting to consider it at the macro level, as the weight of charter schools in terms of students and public expenditure is relatively high compared to neighbouring countries. According to the statistics on "Pupils enrolled in lower-secondary education by program orientation, sex, type of institution and intensity of participation" (Eurostat, 2022a, 2022b, 2022c), in 2020 Spain had the third lowest share of students in lower secondary education studying in public schools (68.27%) and the fourth highest share of students studying in charter schools (27.9%).

One of the population variables related to the phenomenon of social exclusion, which also does not seem to have been studied in detail in the literature on ESL is the fertility rate of adolescent girls. There is evidence of a relationship between higher educational attainment and lower fertility, as well as of the impact of school enrolment on fertility (e.g., Belfield, 2008), and the impact of school enrolment on delayed age at birth (e.g., Ní Bhrolcháin & Beaujouan, 2012). It is therefore expected that there is a relationship between teenage pregnancy and ESL. In this sense, it is also noteworthy that in developed countries (such as the US, Canada, the UK, Northern Ireland, Norway, Hungary), in different time periods and for different population groups, strong evidence has been found of a relationship between increases in compulsory schooling and a lower likelihood of an adolescent student becoming pregnant (see Black et al, 2004; Oreopoulos, 2007; Silles, 2011; Adamecz-Völgyi and Scharle, 2020). From the opposite perspective, a recent paper by Adamecz-Völgyi et al (2021) points out that the 2011 Hungarian education reform, which reduced compulsory schooling, led to an increase in births among what the authors classify as "disadvantaged girls". In the case of Spain, Mora et al (2010) find evidence that an increase in the fertility rate among 15-19 year olds has a positive effect on the dropout rate for girls, but not for boys. Similarly, Bradley and Renzulli (2011) point out that pregnancy or the possibility of becoming a parent is one of the most common reasons for being "pulled out" of the education system. While it may seem that this variable mainly impacts teenage girls, INE data for 2020 (the most recent year examined) show 3,702 births to mothers aged 15-19 with fathers aged under 25, representing 59.3% of all births to teenage girls in this age group. Thus, the teenage fertility rate is likely to have an impact on the early school leaving of boys, especially in terms of generating income in response to new circumstances. (INE, 2022b).

Immigration is another population characteristic considered in the analysis of school failure. Calero et al (2010) find evidence in Spain of a relationship between the immigrant origin of students (first generation) and the likelihood of low PISA scores, which can be assumed to indirectly lead to higher dropout rates. According to O'Hanlon (2016), students belonging to ethnic minorities suffer from higher ESL rates. More specifically, other analyses of the Spanish case have concluded that non-national students are more likely to drop out than nationals, although there are large differences by region (Roca, 2010) and is more pronounced for males (Fernández-Macías et al., 2013). Another issue that has been analysed in relation to ESL is the differences between first and second or third generation immigrants. De Witte et al (2013) note for the case of the United States that studies on the ESL behaviour of second or third generation immigrants do not allow for a clear conclusion to be drawn, with sometimes higher and sometimes lower rates than for first generation immigrants. One factor to be taken into account is that the origin of the immigrant population is very diverse and its distribution in Spain has changed considerably in recent decades. Therefore, it is essential to distinguish their impact on the immigrant population given the differences that may exist in social, economic and cultural terms especially in relation to previous educational level (Fernández-Macías et al, 2013).

Two other aspects of population characteristics included in this analysis are population density and human capital stock. The former is used as a proxy for student-teacher ratios and class size, which are often considered in ESL analyses (e.g., Mora et al, 2010) and cannot be analysed directly in a macro-level study such as this one. On the other hand, Author's own, [2012] point to the existence of a positive relationship between a higher stock of human capital (measured in terms of individuals' years of schooling) and lower ESL, so it can be assumed that those Spanish regions with a predominantly lower educational level in their population will also have higher ESL.

The labour market situation is also fundamental to understanding ESL, as the need to provide for the family is a key factor in the decision not to continue studying (see e.g., Van Praag and Clycq, 2020, for a recent analysis of the Flemish case). Author's own, [2012] argue that higher unemployment can have a positive or negative impact on school failure. For example, higher unemployment could, on the one hand, increase ESL by reducing the expected returns to educational investment and creating demotivation. This effect can be expected to occur when unemployment is concentrated in more skill-intensive jobs. On the other hand, an increase in unemployment could reduce ESL because there is no incentive to drop out of school due to scarce job opportunities. In other words, the opportunity cost of staying in education is low. This situation seems more likely when unemployment is concentrated in low-skilled occupations. Borgna and Struffolino (2017), in a study with Italian data, provide evidence for the latter scenario, pointing out that one of the main pull factors leading to ESL for boys is the availability of employment, especially if it is low-skilled or informal, and in sectors where there is more male concentration. In the case of Spain, Bayón-Calvo, Corrales-Herrero, and De Witte (2020) carried out a comprehensive review of the literature which shows that ESL increases during periods of economic growth, with the opposite effect occurring in times of crisis. In an earlier study, Roca (2010) pointed out that the increase in low-skilled employment since the mid-1990s has had a pull effect on young Spaniards. Given that the impact on ESL seems to vary according to the characteristics of employment, it seems appropriate to disaggregate it by sector of activity. This has been done by Author's own, [2012] find that Spanish regions with a strong construction sector have more ESL, especially among males, because it offers a quick exit with little educational training. On the other hand, the more industrialised regions, where access to employment depends more on qualifications, have less ESL.

Some of the authors already cited point to some gender differences in the impact of employment on school failure. In this sense, Borgna and Struffolino (2017) argue that the opportunity cost of remaining in education is lower for girls than for boys, considering that girls have more difficulties in finding a job. This study, which originally focused on Italy, is relevant to the Spanish context. In 2020 (the last year of the study), the overall unemployment rate for women was 17.43%, with rates of 60.97% for 16-20 year olds and 36.22% for 20-24 year olds (compared with 13.87%, 50.11% and 34.23% respectively for men). Moreover, in the same year, 74% of part-time jobs were by women (INE). In addition, according to the authors, girls could anticipate future discrimination on the labour market, which would keep them in the education system for longer. Thus, girls would wait until they reached a higher level of educational that would allow them to compete better for a given job or to aspire to higher-skilled positions, considering that in some sectors low-skilled positions are traditionally occupied by men. A more recent study by Struffolino and Borgna (2021) shows that among early school leavers, women are less likely to enter the labour market. These arguments are also consistent with Adserà's work for OECD countries (2004), which links high unemployment to a decline in fertility rates and suggests that many women may choose to continue their studies when the employment context is unfavourable.

This literature review provides a comprehensive overview of factors that contribute to early school leaving (ESL) in both international and Spanish contexts. It examines a variety of determinants, including macroeconomic, psychological, demographic, social, family, school and educational aspects. While there is robust empirical support for some determinants, there is limited or no evidence for others. Key variables include education expenditure, school ownership, immigration, population density and human capital. The literature also examines the link between ESL and the labour market, highlighting the impact of employment challenges and gender differences. The lack of consensus on certain variables motivates a deeper exploration of the causes of ESL. In summary, these studies offer a holistic view of ESL, recognising its multifaceted nature and the relevance of different determinants in different contexts.

## 3. The Spanish context

Spain has 17 autonomous communities and 2 autonomous cities, i.e., 19 regions. Throughout Spain's democratic history, there has been a process of political decentralisation, that has also affected the education system. Educational powers are shared between the State and the Autonomous Communities. In any case, the academic path in Spain has the following stages: 1) early childhood education (0-6 years, voluntary); 2) primary education (6-12 years, compulsory); 3) lower secondary education (12-16 years, compulsory); 4) upper secondary education (usually 16-18 years); 5) vocational training (VT): basic VT (minimum 15 years), intermediate VT (minimum 16 years) or higher VT (minimum 18 years); and 6) university education.

According to the Spanish Labour Force Survey (INE, 2022c), although the ESL rate in Spain reached 16% in 2020, there are significant differences between the different Spanish regions. Moreover, these regions have shown an uneven performance of this indicator over the period. Figure 1 shows the ESL rate for the different Spanish regions in 2002 and 2020, based on data from the Spanish Labour Force (INE, 2022c). Only the Basque Country, Asturias, Cantabria, and Madrid met the European 2020 target. These differences support the regional approach in the case of Spain.

'Figure 1 here'

2 shows the evolution of ESL by gender from 2004 to 2020, using data from the mentioned survey. It clearly shows a downward trend since 2004, for both men and women. However, as already pointed out by Casquero and Navarro Gómez (2010), the behaviour varies somewhat depending on the region.

'Figure 2 here'

Gender differences are not surprising given that, girls generally appear to have higher educational aspirations, better classroom behaviour and higher grades than boys (Buchmann et al, 2008; Fabes et al, 2014; Borgna and Struffolino, 2017). Despite this, there are few studies in international literature that focus on gender differences.

## 4. Data and methods

### 4.1. Data

In order to analyse the influence of macroeconomic conditions on the ESL rate in Spain, both from a regional and gender perspective, we use annual data, disaggregated by Spanish regions, obtained from both the National Statistics Institute (INE, 2022a, 2022b, 2022c) and the Ministry of Education and Vocational Training (MEFP, 2022), between 2002 and 2020. The definition of each variable is given in Table 1, and their main descriptive statistics for the whole country are given in Table A. 1 (see appendix). In both tables, the variables are grouped according to the three main macroeconomic areas whose influence we have considered: 1) public expenditure on non-university education, 2) characteristics of the population, 3) characteristics of the labour market.

'Table 1 here'

The descriptive analysis reveals gender differences in the ESL rate, reflecting a 'masculinisation' trend observed in other countries in our context. This gender gap, evident nationally (2), persists consistently across regions (3) over the years. Our aim is to identify unique patterns in the effects of explanatory variables on ESL, stratified by gender.

'Figure 3 here'

An analysis of the evolution of the ESL rate in all regions (3) shows that: 1) there is a decrease in all regions; 2) some regions are reducing the ESL rate more than others, although it depends on the starting value; 3) there are large differences in the ESL rates, so that some regions, such as the Balearic Islands or Andalusia, show comparatively high values, while the Basque Country or Navarre show below-average behaviour throughout the period; 4) the ESL rate is higher for males than females in all years and regions; 5) the trend for males and females is similar throughout the period analysed (parallel lines), although in some regions the behaviour is more erratic and there are occasional differences that break this dynamic. Given the expansion of education in Spain over the last four decades (79.9% of the population had ISCED 0-2 in 1978, compared to 45.2% in 2022; 18.2% had ISCED 3-4 in 1978, compared to 20% in 2022; and 1.8% had ISCED 5-8 in 1978, compared to 34.7% in 2022, according to INE data), it is reasonable to assume that the educational level of parents is also increasing, so that young people in 2020 are more likely to have parents with a higher level of education than those in 2002. Research suggests that parents seek to ensure that their children are educated at least at their own level and to avoid downward educational mobility —see Chevalier et al. (2009) and Lee & Lee (2021) for a literature review. Therefore, a gradual decline in ESL rates is expected over time, especially across different cohorts.

In the appendix we have included some figures showing the evolution of some explanatory variables disaggregated by region (Figure A.1 and Figure A.3) as well as other figures that attempt to approximate the possible bivariate correlations that seem to exist between the ESL rate and the same explanatory variables (Figure A.2 and Figure A.4).

**4.2. Methodology**

We used 19 years of cross-sectional data from the 17 regions of Spain, as shown in Table 1, to conduct our analysis. This longitudinal dataset allowed us to use panel data methodology, which provides greater precision than individual cross-sectional analyses. By using panel data methodology, we were able to control for time-varying unobservable factors, while maintaining homogeneity across regions and capturing inter-regional variation that remains stable over time.

In particular, our dataset lacks individual-level data with region-specific identification for each case, which precludes the hierarchical regional structure necessary for multilevel or mixed-effects estimations. Instead, our data are organised by a hierarchical temporal structure in years from 2002 to 2020, making panel data models a more suitable choice.

The specification of the panel model estimated is as follows:

$$Y_{it} = \beta_0 + \beta_1 X_{1it} + \beta_2 X_{2it} + \beta_3 X_{3it} + \alpha_i + \varepsilon_{it} \qquad (1)$$

Where $i$ refers to the 17 regions, $t$ refers to the 19 years under study, $Y_{it}$ refers to the ESL rate of each region in each year, $\beta_0$ is the constant term, $\beta_1$ is the vector of coefficients of the matrix of explanatory variables referring to the public expenditure of each region in each year ($X_{1it}$), $\beta_2$ is the vector of coefficients of the matrix of explanatory variables referring to the characteristics of the population of each region in each year ($X_{2it}$), $\beta_3$ is the vector of coefficients of the matrix of explanatory variables referring to the labour market in each region in each year ($X_{3it}$), $\alpha_i$ refers to the unobservable component specific to each panel and which is invariant over time, and $\varepsilon_{it}$ refers to the idiosyncratic error term, which varies across regions and over time.

When estimating model (1), we consider that the regions may have their own idiosyncrasies which, being unobservable, cannot be included in the specification as another explanatory variable, so this aspect would be included in the term $\alpha_i$. However, it is likely that this idiosyncrasy is in some way related to the explanatory variables themselves ($X_{1it}$, $X_{2it}$ y $X_{3it}$). For this reason, we consider that the model should be estimated using fixed effects, as this is the way to obtain unbiased and efficient estimators when the correlation between $\alpha_i$ and the explanatory variables is not zero. The Hausman test for fixed versus random effects confirms the need to estimate the panel with fixed effects ($\chi^2_{11}$=76.8; p=0.000).

In the panel model with fixed effects, the possibility of including year fixed effects is also tested using the Wald test ($F_{17,261}$=2.86; p=0.000), which initially indicated the need to include year fixed effects, but as no year was found to be significant in the model, the time fixed effect was finally not included.

In the validation of the model with fixed effects, but without year fixed effects, the existence of serial autocorrelation in the idiosyncratic error term is confirmed by the Wooldridge test ($F_{1,16}$=53.599; p=0.000) —hence the need to include an autocorrelation term in the estimation— (Wooldridge, 2010), as well as heteroscedasticity, using the modified Wald test ($\chi^2_{17}$=92.90; p=0.000) (Greene, 2000, p. 598). Finally, the existence of cross-sectional independence is tested using the Friedman test ($\chi^2$ =21.788; p=0.1502) (Friedman, 1937) and the Pesaran test ($\chi^2$ =1.159; p=0.2465) (Pesaran, 2004).

Therefore, finally, the panel model (1) is estimated using the Driscoll-Kraay approximation, with fixed effects and first order autocorrelation, assuming heteroskedasticity and cross-sectional independence, which yields robust standard errors (Driscoll & Kraay, 1998).

## 5. Results and discussion

Table 2 shows the results obtained in the panel data model estimated for the total population, and Figure 4 does so for males and females. To facilitate comparison, we display the average marginal effects of the Driscoll-Kraay estimation in them.

'Table 2 here'

'Figure 4 here'

All models are globally significant and have a high goodness of fit, with R-squared values above 74%. Moreover, almost all explanatory variables included in the models are significant.

### 5.1. Educational public expenditure on public schools and private charter schools

The first set of variables analysed concerns expenditure on education. On the one hand, public expenditure per student on non-university education is analysed, measured in euros per student, and on the other hand, public expenditure on private education financed by regional governments is measured as the percentage of public funding in private charter schools out of total public expenditure on education.

The results show that neither variable is significant for total ESL and for boys' ESL. However, there is a significant effect for girls, although it is relatively small. According to the model, a 1€-increase in public education expenditure per student reduces the ESL rate for girls by 0.001%. Thus, an increase of 1000€ would lead to a 1% reduction in the dropout rate. The variable expenditure in charter schools shows a larger result. Specifically, a 1% increase in the share of spending on charter schools would reduce the ESL for girls by 0.44% (not significant for males or for the total).

In view of the results obtained, public expenditure seems to counteract some educational push-out factors but contrary to Mora, Escardíbul and Espasa (2010), only in the case of girls. This could be partly explained by the fact that girls are more resilient than boys (see, for example,

Buchmann et al., 2018). Studies such as that by Borgna and Struffolino (2017) support this thesis with Italian data showing that among students with lower grades and whose parents have lower educational level, the propensity to drop out is higher for boys than for girls.

On the other hand, it stands to reason that as public spending increases the attractiveness of education, the pull factors to labour market may weaken. The gender differences found in the empirical analysis suggest that these pull factors may be weaker or less consistent for girls, given the greater employment difficulties that girls face compared to boys in the Spanish labour market.

### 5.2. Population characteristics

All population characteristics are significant and positive in the three models. A 1‰ increase in the fertility rate of girls aged 15-19 is correlated with increases in ESL of 0.66% (total model), 0.76% (males) and 0.57% (females). While teenage fertility serves as a proxy for wealth in Spanish regions, previous model specifications included GDP per capita as a regressor, but it proved insignificant. Therefore, this variable is likely to capture additional socioeconomic and cultural aspects, that deserve further discussion. For girls, pregnancy-related complications such as gestational diabetes, nausea, hypertension, anxiety or postpartum depression are obvious challenges. Although not universal, these problems can interfere with academic pursuits. Although many of these challenges are temporary, the reality after childbirth changes significantly for both genders. Beyond the responsibilities of caring for the child, adapting to this new reality requires additional work. As Bradley and Renzulli (2011) argue, pregnancy is a factor that leads to exclusion from the education system. Surprisingly, contrary to the findings of Mora et al. (2010), the impact is more significant for men than for women. This suggests the dominance of a two-parent family model where the man is the main breadwinner, following the "breadwinner and housewife" model (see, for example, Picchio, 1992). In the context of teenage pregnancy, it appears that men tend to leave school earlier than women, who may make additional efforts to remain in the education system for a longer period. Given the challenges girls face in building a career without qualifications (Borgna and Struffolino, 2017; Struffolino and Borgna, 2021), it's plausible that girls are more likely to return to school after childbirth or during the early years of childrearing. However, their decision to return to school or continue their education may depend on parental support and the perceived value of education in their socioeconomic context.

The results indicate a positive and significant correlation between population density, measured as inhabitants per square kilometre, and ESL. The coefficients are 0.059, 0.061 and 0.058 for the total, men, and women, respectively. Despite Spain experiencing an annual average increase of 0.6 inhabitants per square kilometre over the period analysed, the impact of this variable on ESL appears to be relatively negligible. Nevertheless, as outlined in Section 2, lower population density could be correlated with lower student-teacher ratios and smaller class sizes, factors that are associated with lower ESL rates (Mora et al, 2010). However, it is worth considering that densely populated urban centres may also offer the best employment prospects, making them more attractive to early school leavers entering the labour market.

The variable unskilled population, measured as the percentage of adults aged 25-64 with ISCED 2 or less, is also statistically significant in all three models. A 1% increase in this variable increases ESL by 0.32% for the total, 0.47% for males and 0.17% for females. These findings are consistent with the description of the relationship between human capital and ESL by Author's [2012], who emphasise the influence of the socioeconomic environment on the decision to continue or drop out of school. Students may consider dropping out if individuals in their social context, particularly role models with acceptable or successful employment pathways, have already made this decision. The prospect of an unimpeded long-term career could act as an incentive to leave education (see Van Praag & Clycq, 2020). This could also highlight gender differences, with girls more likely to stay in education if they anticipate greater career challenges in their social environment.

The next set of variables analysed relates to the distribution of the immigrant population by origin. The results obtained show that a 1% increase in European immigration lead to an increase in ESL of 0.72% in the overall model, 0.54% for males and 0.91% for females. In the case of African immigration, the same effect produces an increase of 0.82%, 0.68% and 0.96% respectively, while these values are 0.75%, 0.61% and 0.90% when the increase occurs in Central and South American immigration. It should be noted that a proportion of immigrants aged 18-24 who come to Spain do not have secondary education (higher or lower) and do not decide to continue their studies when they arrive in Spain (Fernández-Macías et al, 2013). Thus, the mere increase in the foreign population in Spain, whose growth rate has increased significantly since the 2000s, would partly explain a direct increase in the ESL indicator.

In general, the results in relation to immigration are consistent with the findings of Roca (2010) and Fernández-Macías et al. (2013), indicating that foreign students are more likely to drop out than their national counterparts. However, it is worth noting that, while Fernández-Macías et al. (2013) found a greater impact on males, our results show the opposite. Given that the sample analysed by the author dates from 2007, the question arises as to whether the changes in Spain over the last decade, such as increased migration flows and shifts in their origins, could explain this discrepancy. As these authors attribute the dropout of non-national minors to difficulties in integrating into the Spanish education system, in would be particularly instructive to look more closely at this aspect in the future. The data obtained represent a first step towards a more comprehensive analysis of this multidimensional phenomenon, which may be linked to peer-effects, age of emigration, generational differences, integration, attention to diversity programmes in different regions and other factors influencing the departure from the educational system or the attraction of the labour market. With regard to the latter, it is conceivable that the increase in job opportunities that do not require high qualifications in the service sector, where some of these young people eventually find employment, may contribute to their decision to drop out.

**5.3. Labour market**

The third block of variables examined relates to the labour market. The results show that the youth unemployment rate is not significant when analysing ESL for males, but it is significant for females and the total population. However, the effect is relatively small. Specifically, a 1% increase in this rate leads to a 0.06% increase in total and female ESL. These findings are consistent with the research by Authors [2012], which highlights that rising unemployment can contribute to dropout rates by reducing the expected returns to education and reducing incentives to remain in the education system for longer. As noted in section 2, a lack of incentives and demotivation to continue studying seems more likely when unemployment is concentrated in skilled jobs. This may explain why the effect is mainly observed for girls, as their additional challenges in entering the labour market compared to boys lead them to pursue higher levels of education in order to access higher-skilled jobs (Borgna & Struffolino, 2017).

The last two variables are related to employment sectors, specifically the employment rate in industry and construction. The latter is significant in all three models. A 1% increase in the share of those employed in this sector leads to an increase in total ESL, male ESL and female ESL by 0.64%, 0.77%, and 0.52%, respectively. These results are consistent with Author's own analysis [2012], which suggests that the construction sector, with its demand for low-skilled labour (Observatorio Industrial de la Construcción, 2017) and relatively high wages, acts as an incentive for early school leaving. It allows young people to enter the labour market quickly without the need for further education. The higher impact on boys can be attributed to the masculinisation of the sector, with almost 90% of workers being men, according to INE (2022c). Given that women make up only 10% of the sector, and many of them are highly qualified (Observatorio Industrial de la Construcción, 2017), the coefficients for girls may be explained by their tendency to leave school when their male counterparts secure well-paid jobs in the sector.

In the case of the industrial sector, a 1% increase leads to a 0.47% increase in ESL for men, but is not significant in the total or in the pattern for women. The explanation for men may be similar to that for construction, given that a significant proportion of the jobs created in this sector do not require tertiary qualifications. The lack of impact for girls may be due to the small proportion of the total number of people employed in this sector (7.8% in 2021, according to INE, 2022c). Therefore, also in this case, the results are in line with the work of Borgna and Struffolino (2017), with girls staying longer in the education system, given the additional difficulties in obtaining a low-skilled job and the expectation of a worse employment path if they do not achieve a higher level of education.

## 6. Conclusion

Over the last two decades, ESL rate in Spain has decreased significantly, although it has not reached the European target of not exceeding 10%. The new EU horizon states that ESL should not exceed 9% by 2030, so the country is facing a new challenge that has led us to analyse in depth the causes that explain this circumstance, in order to guide policymakers and Spanish education authorities. As in other Southern European countries this phenomenon has been influenced by some labour market and population characteristics. We use panel data to examine which determinants have shaped these trends in Spain since 2002.

Our methodological contribution is therefore threefold: 1) longitudinal analysis, 2) with a focus on aggregate characteristics at the regional level, and 3) with a gender perspective. For both genders, the influence of the share of unskilled population, the fertility rate of younger women, the population density, the immigrant population (by origin) and the weight of the construction sector in the region on the decision to drop out of school stands out, although to a different extent. The weight of the industrial sector is also significant for ESL, but only for boys. Surprisingly, youth unemployment has only a small but significant impact on female ESL. Finally, policy decisions to increase public spending per student and to increase funding for charter schools would slightly reduce ESL, but only for girls.

Some of these results are consistent with the literature, although there are some novel findings. In this way, we hope to better target socio-educational policies in order to address the problem in a comprehensive manner, paying special attention to the characteristics of the population in Spain, although some recommendations can be extended to other southern European countries at the same crossroads.

With regard to the characteristics of the population, a particularly interesting result is the impact of teenage pregnancy on ESL, which is greater than that of other types of variables traditionally considered in this type of study, such as the level of educational expenditure. This suggests that effective sex education in schools and support in making prudent life choices could have a decisive impact on ESL. The relevance of the fertility variable opens an interesting topic for future research to guide the educational and social authorities in Spain in order to prevent ESL among young adolescents, as has been done in other countries, such as France.

Another notable finding is that immigration is a complex phenomenon influenced by various factors, including social and integration policies in the Spanish regions. The discussion highlights areas for future research, such as exploring the age of immigration, generational differences and other demographic factors influencing exclusion from the education system. The importance of considering the gender of immigrants in future analyses in emphasised, as regional differences in the proportion of men and women, can lead to nuanced insights tailored to the specific characteristics of each region.

In the context of the Spanish labour market, the gender differences found provide food for thought. Given that the motivation that keeps girls in the education system may be undemined when unemployment rises, particularly when it is concentrated in higher-skilled jobs, policies aimed at reducing youth unemployment and creating skilled jobs may be particularly relevant to reducing female ESL. While employment issues are beyond the scope of education policy, one issue that can be strengthened during compulsory education is career guidance and the provision of timely and expert information to students and families about different employment options and their implications. This could be important in reducing early school leaving, especially among boys, given the availability of unskilled employment in some sectors.

Finally, although the scope of public policy does not allow for immediate radical changes in certain characteristics of the population and the labour market, it does seem possible to provide certain incentives to keep young people in education longer. At a time when the social and labour market context is changing at an accelerated pace, recommendations to reduce ESL need to be practical and take into account population-related aspects (such as ethnic minorities, exiles or people with disabilities) that may not have been sufficiently taken into account in Spain so far.

Our work is not without limitations. Indeed, given its macro-level focus, the implications may be less straightforward to apply compared to micro-level studies. However, our analysis allows for policy recommendations at the national level for the Spanish government and at the regional level for the Autonomous Communities. Another limitation is the lack of data disaggregation by autonomous communities over time, which has hindered our ability to work with other variables, such as the average educational level of the parents of early school leavers, and to analyse the data by gender, socioeconomic status or country of origin among the immigrant population.

In future research, it will be important to carry out micro-level analyses to take account of these and other new aspects, such as the digital divide. This divide, which has been revealed by the pandemic in a global context and not only in Spain, has generated a profound social and educational change, increasing the differences in access to and use of new technologies among students at all levels of pre-university education, which in turn could affect ESL.

# Appendix

Table A. 1. Principal descriptive statistics of the variables for Spain

| Variable | | Mean | Std. dev. | Min | Max | Observations |
|---|---|---|---|---|---|---|
| *Variables of interest* | | | | | | |
| Early school leaving | overall | 23.96 | 8.54 | 6.51 | 43.58 | N = 323 |
| | between | | 6.20 | 11.66 | 33.20 | n = 17 |
| | within | | 6.05 | 10.95 | 35.87 | T = 19 |
| Male early school leaving | overall | 28.92 | 10.20 | 8.49 | 52.02 | N = 323 |
| | between | | 7.15 | 14.61 | 39.27 | n = 17 |
| | within | | 7.48 | 14.62 | 43.92 | T = 19 |
| Female early school leaving | overall | 18.76 | 7.24 | 4.27 | 37.89 | N = 323 |
| | between | | 5.28 | 8.57 | 26.79 | n = 17 |
| | within | | 5.11 | 4.37 | 30.38 | T = 19 |
| *Variables related to educational expenditure* | | | | | | |
| Educational expenditure | overall | 5928.19 | 1786.05 | 1416.95 | 13548.90 | N = 353 |
| | between | | 1544.63 | 4294.48 | 11106.06 | n = 17 |
| | within | | 958.06 | 2661.07 | 8371.04 | T = 20.8 |
| Charter expenditure | overall | 17.50 | 5.95 | 6.35 | 31.29 | N = 353 |
| | between | | 5.85 | 8.92 | 29.78 | n = 17 |
| | within | | 1.76 | 10.01 | 23.03 | T = 20.8 |
| *Variables related to population characteristics* | | | | | | |
| Fertility rates for girls aged 15-19 | overall | 8.62 | 3.51 | 2.66 | 21.33 | N = 374 |
| | between | | 2.59 | 4.88 | 15.09 | n = 17 |
| | within | | 2.45 | 1.25 | 15.51 | T = 22 |
| Population density | overall | 159.63 | 175.31 | 21.72 | 844.53 | N = 374 |
| | between | | 179.60 | 24.79 | 766.16 | n = 17 |
| | within | | 17.17 | 34.40 | 238.00 | T = 22 |
| Unskilled population | overall | 46.59 | 9.53 | 24.07 | 68.26 | N = 323 |
| | between | | 7.50 | 32.93 | 60.67 | n = 17 |
| | within | | 6.14 | 33.83 | 61.17 | T = 19 |
| European immigration | overall | 42.21 | 12.13 | 14.95 | 77.37 | N = 374 |
| | between | | 10.24 | 25.23 | 60.79 | n = 17 |
| | within | | 6.93 | 20.60 | 62.56 | T = 22 |
| African immigration | overall | 20.71 | 9.98 | 6.83 | 59.74 | N = 374 |
| | between | | 8.83 | 10.49 | 38.84 | n = 17 |
| | within | | 5.11 | 8.50 | 45.22 | T = 22 |
| Central and South American immigration | overall | 30.33 | 11.19 | 7.40 | 57.70 | N = 374 |
| | between | | 8.77 | 17.32 | 42.84 | n = 17 |
| | within | | 7.27 | 7.33 | 48.34 | T = 22 |
| *Variables related to labour market* | | | | | | |
| Youth unemployment rate | overall | 34.45 | 13.64 | 9.63 | 66.67 | N = 340 |
| | between | | 5.05 | 28.31 | 44.20 | n = 17 |
| | within | | 12.73 | 10.98 | 62.17 | T = 20 |
| Employed in agriculture sector | overall | 5.69 | 3.73 | 0.10 | 17.77 | N = 391 |
| | between | | 3.52 | 0.49 | 12.01 | n = 17 |
| | within | | 1.47 | 2.47 | 14.14 | T = 23 |
| Employed in industry sector | overall | 16.84 | 6.48 | 4.10 | 32.86 | N = 391 |
| | between | | 6.26 | 5.66 | 27.02 | n = 17 |
| | within | | 2.24 | 12.17 | 24.16 | T = 23 |
| Employed in construction sector | overall | 9.34 | 3.19 | 4.38 | 16.95 | N = 391 |
| | between | | 1.30 | 7.35 | 12.13 | n = 17 |
| | within | | 2.93 | 4.26 | 16.65 | T = 23 |
| Employed in service sector | overall | 68.13 | 8.42 | 47.29 | 87.95 | N = 391 |
| | between | | 6.59 | 58.00 | 81.07 | n = 17 |
| | within | | 5.47 | 56.19 | 76.55 | T = 23 |

Note 1: N = total number of observations; n = number of regions; T = number of observed years

Note 2: The mean for the whole sample is presented as well as other measures of dispersion, such as standard deviation and maximum and minimum values, both for the whole sample (overall), between regions (between) and between years (within).

Figure A.1: Evolution of fertility and immigration rates by region

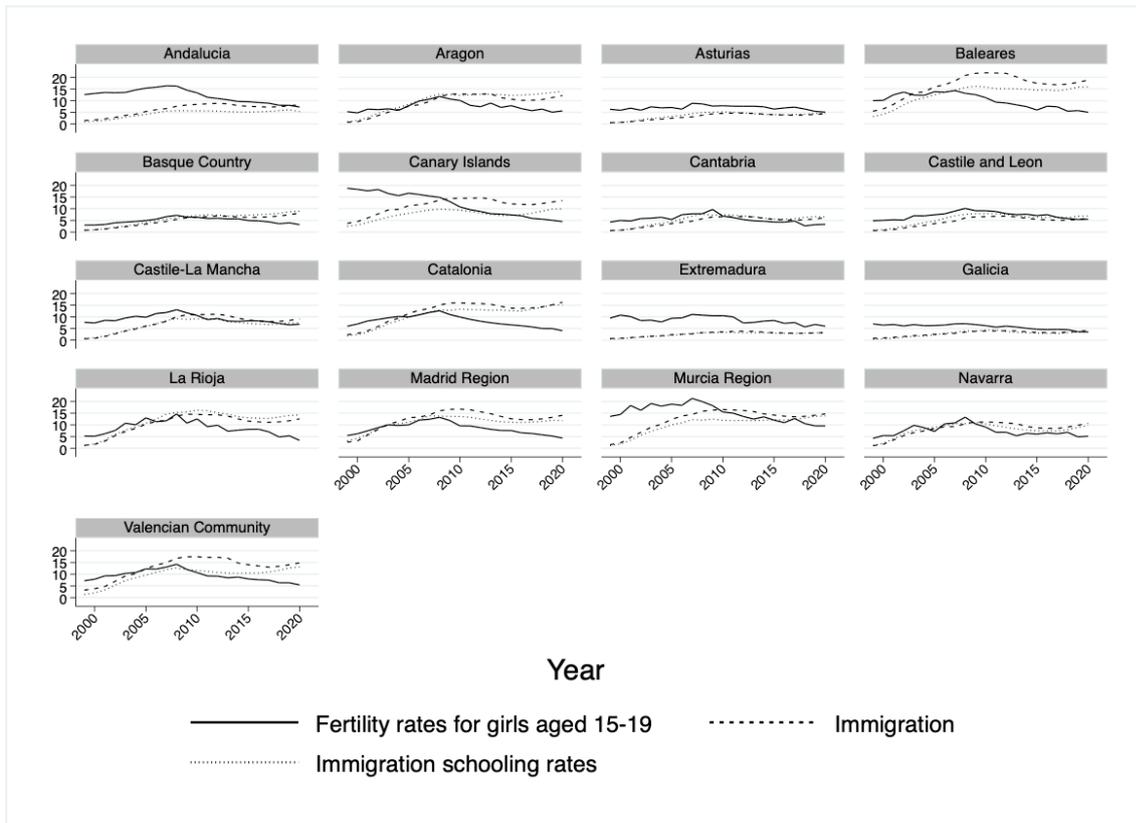

*Note:* A decreasing trend in the fertility rate can be observed in the different regions, as well as an increasing trend in terms of immigration.

*Source:* Adapted from 2022 Demographic Indicators, by National Statistics Institute (INE, 2022b).

Figure A.2: Correlation between ESL, fertility rate and immigration for the pooled data between 2002 and 2020

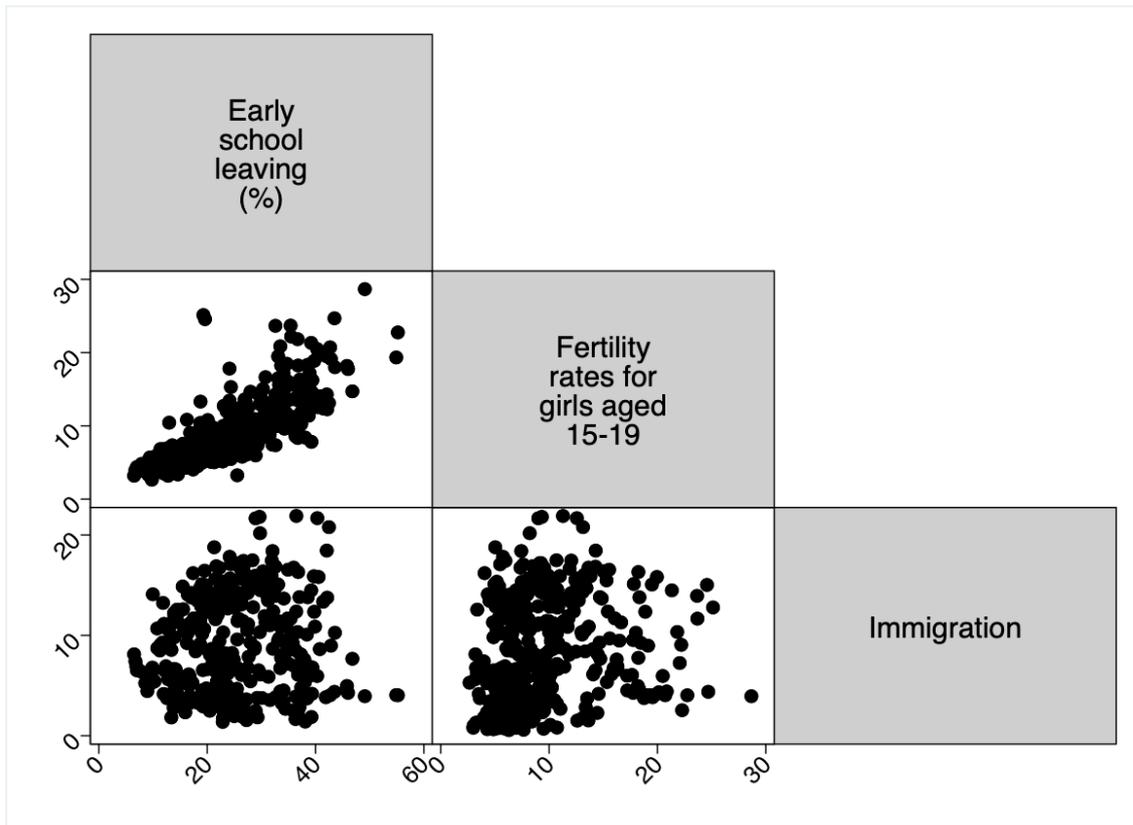

*Note:* Preliminary correlation analysis shows a positive relationship between ESL and fertility rate (increases in fertility rate lead to higher dropout). However, there is no clear relationship between ESL and immigration data, which requires further analysis and will eventually result in separating immigrants according to country of origin.

*Source:* Adapted from 2022 Continuous Census Statistics, Demographic Indicators, and Spanish Labour Force Survey, by National Statistics Institute (INE, 2022a; 2022b; 2022c).

Figure A.3: Employed distributed in activity sectors across the period 2002 and 2020 by regions.

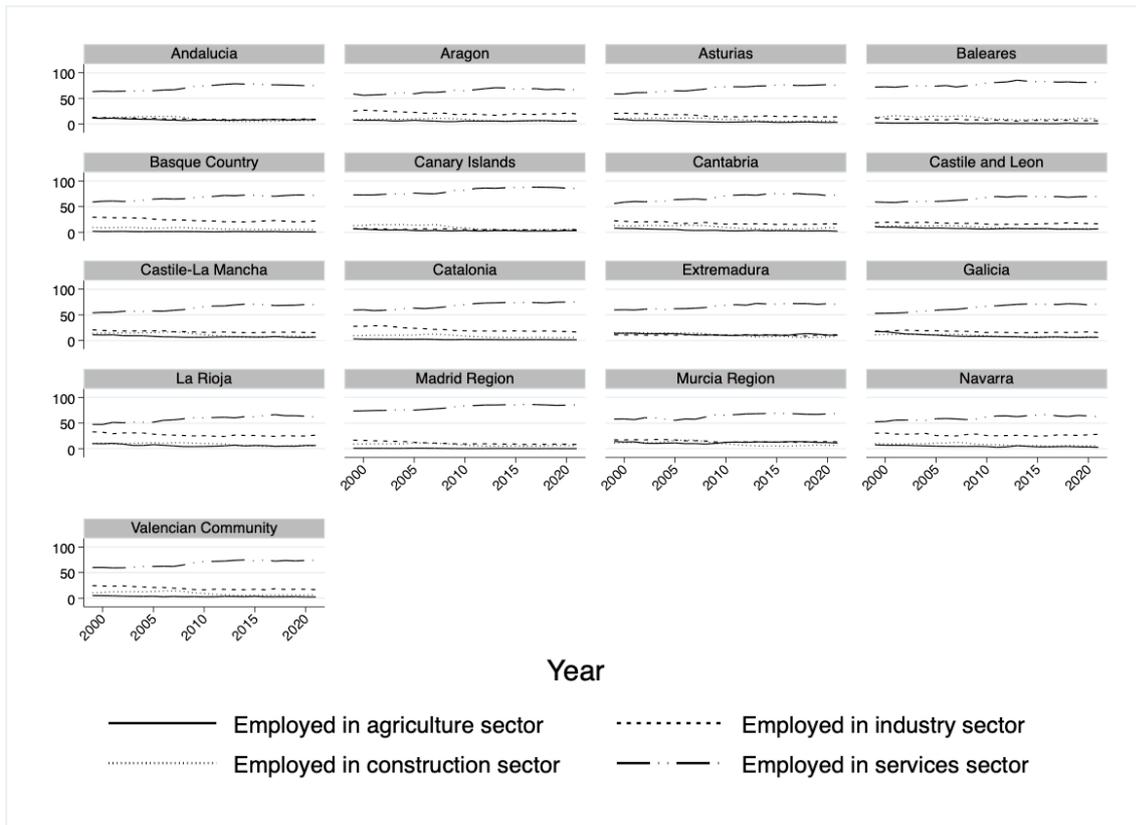

*Note:* It can be seen that the service sector is the most important sector in our country, with a significant percentage of the population employed in this sector compared to the rest.

*Source:* Adapted from 2022 Spanish Labour Force Survey, by National Statistics Institute (INE, 2022c).

Figure A.45: Correlation between ESL and employment by sector of activity for the pooled data between 2002 and 2020 by region

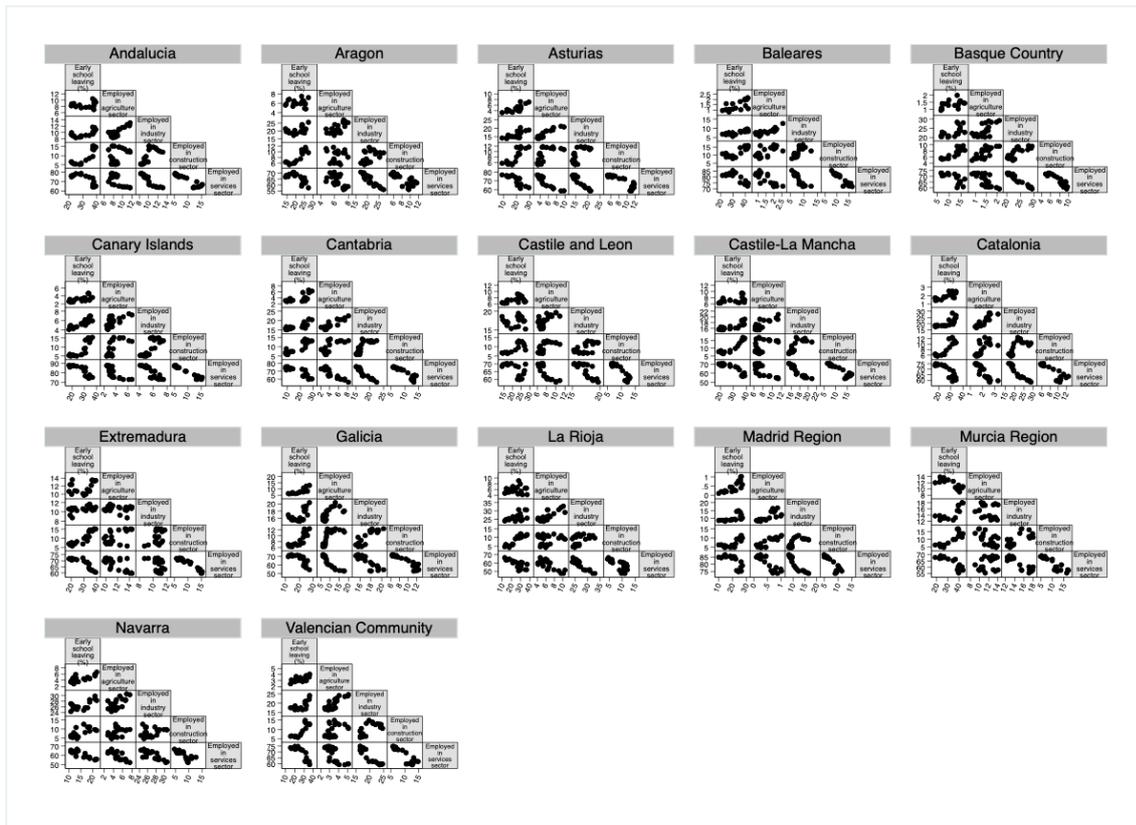

*Note:* In general, there is a positive correlation between ESL and occupation in all sectors of activity, except in the services sector, where the correlation appears to be negative.

*Source:* Adapted from 2022 Continuous Census Statistics, Demographic Indicators, and Spanish Labour Force Survey, by National Statistics Institute (INE, 2022a; 2022b; 2022c).

Figure 1. ESL in Spanish regions, 2020 and 2002

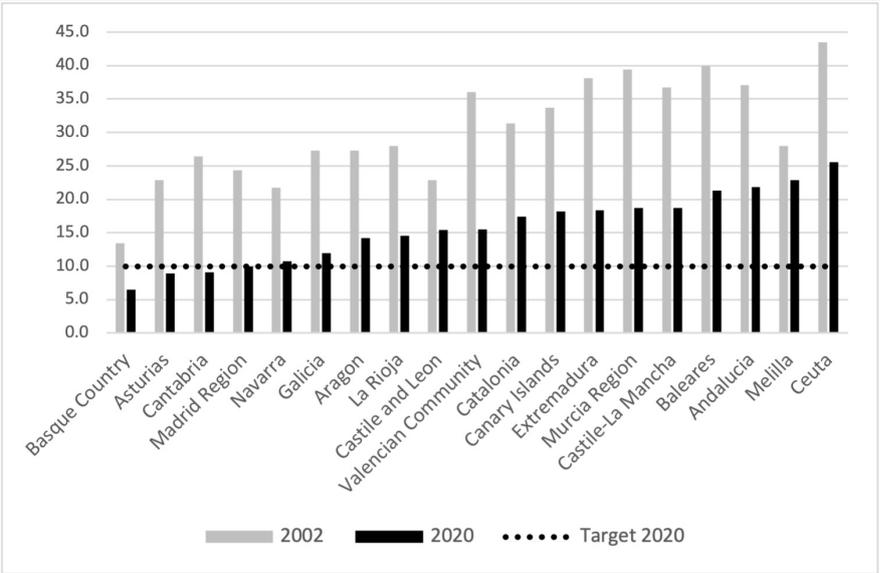

Source: Spanish Labour Force Survey

Figure 2. Evolution of ESL in Spain, by gender, 2004-2020

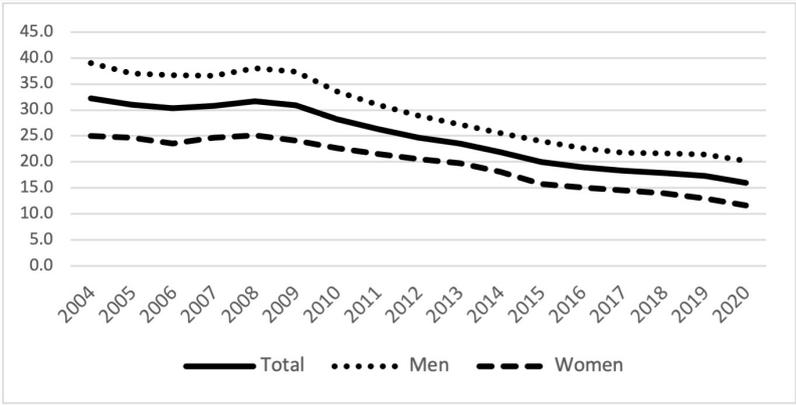

Source: Spanish Labour Force Survey

Figure 3. Evolution of ESL in Spain by gender across the period 2002-2020

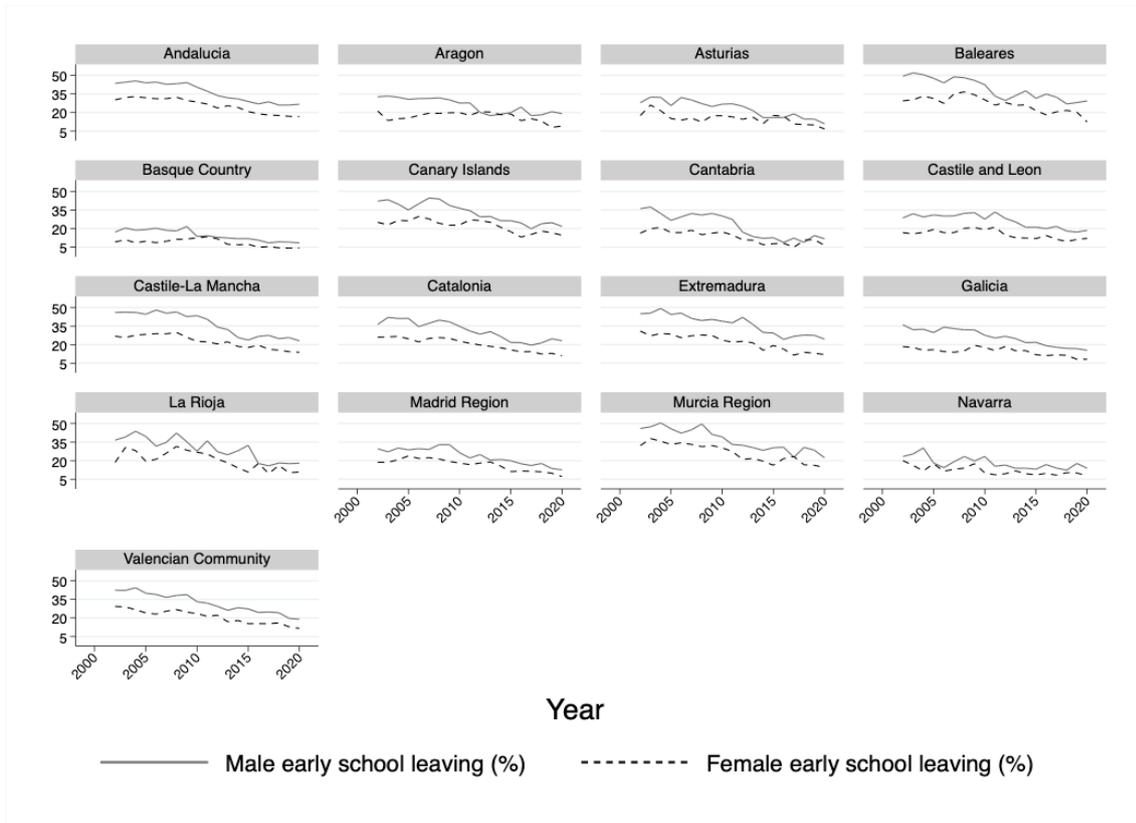

Figure 4. Panel Regressions with Driscoll-Kraay standard errors – Marginal effects with 95% confidence intervals for male and female early school leaving rates

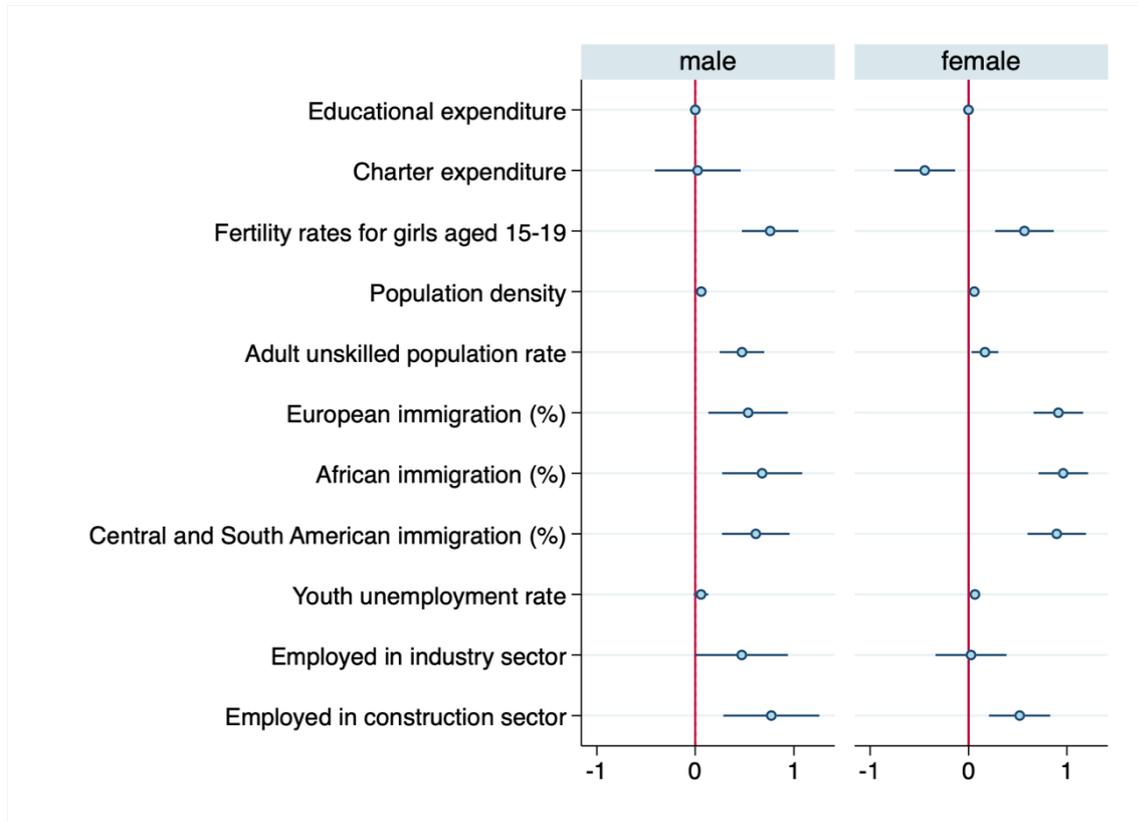

Note 1: Goodness of fit measures for males panel-estimations: $F(11,16) = 1959.54$, Prob > F =0.000, within R-squared= 0.8480; Goodness of fit measures for females panel-estimations: $F(11,16) = 624.44$, Prob > F =0.000, within R-squared= 0.7446

Note 2: The only non-statistically significant variable in the estimation for female ESL pertains to employees in the industrial sector. In the case of male ESL, the non-statistically significant variables are educational expenditure, charter expenditure, and the youth unemployment rate.

Source: Author's own elaboration

Table 1: Variables, definitions and source

| Variable | Definition | Source |
| --- | --- | --- |
| *Variables of interest* | | |
| Early school leaving | Proportion of 18-24 year olds, who had completed at most lower secondary education (ISCED 2) and are not in further education or training | INE - Spanish Labour Force Survey (INE, 2022c) |
| Male early school leaving | Proportion of 18-24 year old males, who had completed at most lower secondary education (ISCED 2) and are not in further education or training | INE - Spanish Labour Force Survey (INE, 2022c) |
| Female early school leaving | Proportion of 18-24 year old females, who had completed at most lower secondary education (ISCED 2) and are not in further education or training | INE - Spanish Labour Force Survey (INE, 2022c) |
| *Variables related to educational expenditure* | | |
| Educational expenditure | Public expenditure per student on non-university education (€/student) | MEPF – EDUCAbase (MEPF, 2022) |
| Charter expenditure | Public expenditure on charter education (percentage of total educational public expenditure) | MEPF – EDUCAbase (MEPF, 2022) |
| *Variables related to population characteristics* | | |
| Fertility rates for girls aged 15-19 | Fertility rates (x1000 women) for girls aged 15-19 years | INE - Demographic indicators (INE, 2022b) |
| Population density | Number of inhabitants per square kilometer | INE - Demographic indicators (INE, 2022b) |
| Unskilled population | Rate of unskilled adult population (% of adults, aged 25-64, with ISCED 2 at most) | INE - Spanish Labour Force Survey (INE, 2022c) |
| European immigration | Share of European immigration in total immigration | INE - Continuous census statistics (INE, 2022a) |
| African immigration | Share of African immigration in total immigration | INE - Continuous census statistics (INE, 2022a) |
| Central and South American immigration | Share of Central American and South American immigration in total immigration | INE - Continuous census statistics (INE, 2022a) |
| *Variables related to labour market* | | |
| Youth unemployment rate | Unemployment rate among 16-24 years old | INE - Spanish Labour Force Survey (INE, 2022c) |
| Employed in agriculture sector | Employment rate in the agriculture sector | INE - Spanish Labour Force Survey (INE, 2022c) |
| Employed in industry sector | Employment rate in the industry sector | INE - Spanish Labour Force Survey (INE, 2022c) |
| Employed in construction sector | Employment rate in the construction sector | INE - Spanish Labour Force Survey (INE, 2022c) |
| Employed in service sector | Employment rate in the service sector | INE - Spanish Labour Force Survey (INE, 2022c) |

Table 2. Regression with Driscoll-Kraay standard errors - Total early school leaving rate

| TOTAL ESL | Coefficient | | [95% conf. interval] | |
| --- | --- | --- | --- | --- |
| Educational expenditure | -0.000 | | -0.001 | 0.001 |
| Charter expenditure | -0.199 | | -0.515 | 0.117 |
| Fertility rates for girls aged 15-19 | 0.664 | *** | 0.436 | 0.893 |
| Population density | 0.059 | *** | 0.037 | 0.081 |
| Unskilled population | 0.321 | *** | 0.154 | 0.487 |
| European immigration (%) | 0.720 | *** | 0.490 | 0.950 |
| African immigration (%) | 0.815 | *** | 0.611 | 1.019 |
| Central and South American immigration (%) | 0.754 | *** | 0.546 | 0.963 |
| Youth unemployment rate | 0.060 | ** | 0.008 | 0.111 |
| Employed in industry sector | 0.255 | | -0.052 | 0.561 |
| Employed in construction sector | 0.641 | *** | 0.333 | 0.949 |
| Constant | -84.200 | *** | -100.221 | -68.178 |
| F(11,16) = 768.48 | Prob > F =0.000 | | within R-squared= 0.8741 | |

Note: Significant at 1% ***; 5% **; 10% *.

Source: Author's own elaboration